# IS A GLOBAL VIRTUAL CURRENCY WITH UNIVERSAL ACCEPTANCE FEASIBLE?

Sowmyan Jegatheesan, Sabbir Ahmed, Austin Chamney And Nour El-kadri*

*University Of Ottawa, Canada*

**ABSTRACT**

As digital goods and services become an integral part of modern day society, the demand for a standardized and ubiquitous form of digital currency increases. And it is not just about digital goods; the adoption of electronic and mobile commerce has not reached its expected level at all parts of the globe as expected. One of the main reasons behind that is the lack of a universal digital as well as virtual currency. Many countries in the world have failed to realize the potential of e-commerce, let alone m-commerce, because of rigid financial regulations and apparent disorientation & gap between monetary stakeholders across borders and continents. Digital currency which is internet-based, non-banks issued and circulated within a certain range of networks has brought a significant impact on the development of e-commerce. The research and analysis of this paper would focus on the feasibility of the operation of a digital currency and its economic implications.

# INTRODUCTION

The idea is to have risk free transfer of currency and the value of it across social and economic barriers. This involves multiple stakeholders, governments, organizations, banks and users. The concept of virtual currency can certainly influence e-commerce but will that be possible? That's the question we are trying to answer. We are trying to do a feasibility study through a survey and literature which is available. Various factors are taken into account during the research that includes business and economic interests and well as implementation and security features of the currency. Though this paper's focus is to identify the causes there are no solutions proposed. Some of the important questions that are circling around the possibility of such a feature are added to the survey and research after a though analysis of the current market situation and real life difficulties in terms of usage, influence and perception are also taken into account. This research is aimed to be a comprehensive and a thorough analysis of the practical situation. The most important questions that need to be understood on what is the purpose of this currency and what is the real benefit of this currency? Will this be used or adopted widely? Who will be the real beneficiary? What are the existing problems in the system? These questions form the basis for the research and the study.

# LITERATURE REVIEW

Our feasibility study is focussed to fulfill the gaps and questions raised by many researchers in the virtual currency and ecommerce markets. There are community currencies, digital wallets and e-currencies that have evolved over the period of time, but the there is no comprehensive study on the feasibility of these currencies and their role in enhancing ecommerce and online transactions. This study becomes significant and gives an insight about the existing barriers, limitation that the virtual currencies would have and also the benefits that they might bring. Also the study analyzes the implementation and regulatory issues so that future currency systems should take a note of them and find solutions for it. Since the development of Internet virtual currency would become a new form of payment. But the multi-currency coexistence and inconvertibility of the current virtual money initiated the dispute on whether the virtual currency should be unified or whether such project is feasible (Peng, Niu and Wu, 2009). The prime considerations of virtual currency and economy along with money supply and inter-conversion hold the key in the currencies effective usage. The money supply model and its impact are essential for the feasibility of the virtual currency. The question of currency reserves and benefits are analysed as well (Ouyang and Zhang, 2009). The efforts and expenses in operating the printed currency and the measures taken for the security, maintenance of the current currency circulation are all an anachronism in the information age. Some newly developed system of cryptographic protocols realizes digital currency so that electronic payments can be done without online authorization or pre-registration. The user's electronic wallet contains digital currency, and can be used both at physical and virtual (such as Internet) points of sale (Mjolsnes and Michelsen, 1997). Social networking and e-commerce sites online services can issue scrip to users as a representation of positive value in their dealings with those users. Like cash in a bank account, this currency may be withdrawn and exchanged with a third party. By holding the currency of that issuer, this third party (who may not have had any previous contact with the issuer) may now access resources with whichever privileges are afforded to those holding the issuer's scrip (Hansen, 2010). Virtual money is the basis of building the value exchanging platform between virtual economy and real economy. It has greatly promoted the e-commerce updating and will be an important part of advanced e-commerce. The virtual money can also help to set up a more unified network market economy. So the world should pay close attention to and encourage the development of virtual money (Chen and Wu, 2009).Though virtual currency only has partial function of real currency currently, there have been many problems drawing extensive concern. Some believe virtual currency destroys financial policy and affects currency circulation in reality. As a result, it may cause inflation, therefore virtual currency should be regulated by the central bank; Virtual currency won't interfere with the operation of real currency, instead it may boost the development of IT industry and e-business effectively. Governmental interference will only put an end to such an emerging industry. As the fierce argument goes on, there's still no in-depth research on this regard (Wu, Peng and Zhu, 2009). The emergence of e-commerce has created new financial needs that in many cases cannot be effectively fulfilled by the traditional payment systems. Recognizing this, virtually all interested parties are exploring various types of electronic payment system and issues surrounding electronic payment system and digital currency (Singh, 2009). Currency scale and various approaches to monetary governance need to be, identifying a number of key limitations with previous approaches and highlighting the need for a modified conceptual and theoretical framework for exploring the potential of small scale currency institutions to allow greater participatory monetary decision-making (Jones, 2011).

# SURVEY AND RESPONDENTS

The survey had respondents that included Students, Payment processing Industry professionals, Economists, CEOs, E-commerce Experts and consultants and Business Community. The limitations of the survey were that the number of people we covered across the spectrum from diverse background was good but due to privacy concerns we did not measure the age metrics for the survey. The survey has helped us to understand and analyze the trends in the total research. The survey was conducted online and the survey respondents where all regular users of internet and online shoppers. The survey respondents were selected to be online shoppers because their experience with existing payment methods is essential for the feasibility study. 66 people took the survey.

## RESEARCH QUESTIONS

The research revolves around different questions which we think would determine the feasibility of the digital currency. It could be analysed by some key questions answered among stakeholders. This research questions would introduce the concept of social support as a social effect of community currencies and explores different ways of measuring it (Nakazato and Hiramoto, 2012).

**Importance of a Global Currency**

The importance of having a Global currency would be most important aspect of getting the e-commerce revolution to nook and corner of the world. This specific aspect plays a key role in terms of usability, adaptability of users to such a currency. The key question still would remain whether the virtual currency which may be adopted universally would completely replace the physical present day currencies of different countries or would be used in the sidelines of the existing currency. The research and analysis in the past suggests that there were always dominating currencies in the world in the past as well and at present the prime currency being the United States Dollar (Schenk, 2009). The important factor of such a currency has to be understood. The lack of universal currency for the e-commerce transactions directly reflect in the volume of cross border trade as such transactions end up incurring huge money as foreign Exchange costs, processing fees and commissions. So this ultimately increases the cost of the trade and also the volatility in the international market and brittle financial situation makes buyers and sellers unable to fix the prices or make proper estimates of costs. If such a global currency is brought out who would manage the currency, what would be the value of the currency are some of the key questions that arise. Money being the most important factor and the real purpose of any business or commercial activity the reliability of such a currency is very essential if its needs to get a wider acceptance. The increasing global issues and disagreements with different countries over different social, political issues across the globe makes it very complicated to come up with a global currency for the whole world. The story of Euro is a very good example of such a currency but the consistent financial crisis since its inception has made researchers less aware of the real benefits of euro as the world and the euro zone have been facing a recession and the worst financial crisis which lead to severe financial down grading of various financial institutions and the countries themselves. To get wider acceptance the companies, countries and different stakeholders should see a real benefit for everyone and it should be an absolute win-win for all the involved parties. This really seems unlikely looking at the history of world politics and building consensus on important issues. But the mature understanding is that one currency for all your transaction which we could manage it digitally would be a great feature and best thing human beings can create themselves. But it is inherently complex in the building and also in the operational level. The present direction of global economic change signifies the need for the development of alternate modes of currency as a medium of exchange that can match the fast paced, computerized economy. The future is thus in the use of handheld electronic 'Currency (Cash) Pads' which will replace paper currency for many day-to-day transactions (Sharma, 2005).

**Printing of Currency**

Printing of the present day currency has reduced to a large extent and the volumes of currencies that are printed have not increased as much. Though there are many reasons for this still we know that the world is moving towards digitalization and people are increasingly using payments services like Credit/debit cards, mobile payment systems and online transfers. So this seems a clear indication that the world is moving towards paperless economy and the global currency would certainly be a product of this process in the near future. More over the cost of printing and managing the printed currency has become huge in the recent years because of multiple reasons that include the life of the printed currency going less and security of the currency notes printed (FRB: How much does it cost to produce currency and coin?. 2013). The counterfeit currencies for political and economic gains reasons have added to the trouble of the printed currencies. For two hundred and sixty years the US federal government has claimed that the most democratic money is a scarce form of money. This claim is built off the notion that an abundant supply of money would threaten class relations (the rights of private property) and ultimately the free flow of commerce (capitalist exchange) (Wainwright, 2012).

**Biometrics**

Biometrics being the most important mode of security, identification and authentication in recent times and the enhancement of the same in the operational level was also a key research question we tried to understand. The user comfort in the popular biometric systems is also analysed. It is seen as a privacy concern by many people and they don't wish to share their personal information that would be transmitted across different systems. The invention of new biometric methods that includes Iris, Fingerprint, palm Scanning and the Sweat Gland technologies have proved more reliable and stress free for people who need not carry anything as they themselves can authenticate them. The practical implementation for payments using the biometrics has kicked off in certain areas in the US and has had considerable success but the logistics of the total exercise needs to be understood (Johnson, 2012). We focus on this part to understand whether the global virtual currency would be operational in a biometric mode as we can just travel around the world and make payments with our finger prints or a palm scan which would be the most convenient and secure way of making it.

**Mobile payment system and Mobile operators**

The using of mobile phones as the mode or method of payment is one other operational activity which we tried to analyze. The popularity of NFC technology and its applications have paved way for such a service to the Global cur-rency. It has had practical implementation with payment providers in the US and has been quite successful in the trials (Webb, 2013).The wider adoption is yet to be known and there are various constraints in this particular technology which makes its mandatory for users to have a device that has NFC in it. The NFC has had no great upgrade because till now the iPhone has had no facility for this technology and this has been a great barrier considering the fact that iPhone tops the Smartphone market and the NFC technology has lost considerable user base in terms of both operational and adaptability because of lack of support for the technology from Apple. The other question which circles this is the mobile operators being the billers for the user payments and services. Since a successful model like I-mode which had a huge success in Japan eventual failed in the US. Though there were multiple factors in this the prime factor is the consideration that the mobile operators would bill the users and the users trust level to allow mobile operator top handle the money for them in place of the bank.

**Foreign exchange, the cost and transactions**

The main problem having different currencies is that we need to find the exchange rates on a real time basis and pay a huge amount during the money conversion charges to different service provides. This includes banks, Foreign exchange agents and Payment processing companies. Be it buying things online during cross border transactions or using services that allow sending payment online there is a huge amount of services charges paid. This includes charges for the services, then a fraction of the amount as a commission and sometimes a huge premium in the exchange rate also. This happens across the world and this certainly is huge cost thinking about the overall volume of the transactions that arise. Even day to day banking services that involves cross border transactions or foreign currency transactions are charged at least 3 to 4% extra as charges by payment service providers like VISA, MasterCard etc. These makes up a huge cost for e-commerce transactions and finally make the service or the transaction costly (MasterCard Interchange Rates & Fees | MasterCard. 2013). To solve this there should be a global currency that there is no need for exchange rate. Also we wanted to research on the redemption or valuation of the global currency. How this currency can be valued or sold initially, how to give proper valuation to the currency in a longer run.

**Economic implication because of digital currency**

The global currency when it would be used across the borders there would be serious economic consideration and questions that we need to understand and answer. The concerns would include the volume that is to be circulated that also includes the creation of demand. There are various other factors like GDP, Inflation which needs to be answered with this global currency. What would be the stake or role of each and every country in the total process in terms of economic activity would be a serious question. The management of such a currency would be with a global body that has to be accepted by all the central banks and the roles of each and every central banks and countries along with business and commercial counterparts would be inherently complex. So this is again a complex constraint in the effective operation of any such global currency. But alternative ways have to be found as there is a desperate need in this hour of cyber and digital trading era. The monetary stability, financial sector development and a country's general level of economic development are all positively related to both the likelihood of a country hosting an alternative currency as well to the number of alternative currencies a country is hosting (Pfajfar, Sgro and Wagner, 2012).

**Involvement of government and financial regulators**

The involvement, support of regulatory bodies in any country is very essential in the operation of this currency. Different countries follow different systems and banking methods. A line has to be drawn and the common principles of banking and effective regulations have to be put forward for such a currency to operate to its full tendency and to achieve all the benefits. If the currency loses the support for the regulatory bodies then there is no way that is going to be effectively used by the users. The system should be very well designed to take out the flaws in the existing system that includes tax evasion, money laundering and also give accountability for all the money in the system thereby adding value to the existing system so that it gets wider acceptance even though some governments have different opinions and policies. There is always a way to come forward and plan the regulations effectively to seal the system. This certainly seems possible with a currency that is global and fully digital.

**Cost involved in setup and transactions for e-commerce websites**

There is a huge setup cost that is involved for e-commerce websites when they try to accept payment online and this is problematic both for the buyers and sellers in the transaction. Though there is need for a cost on every transaction the amount that's being charged by the gateway providers, payment processors are way too high for small and medium businesses that this impacts the way e-commerce is perceived. To change this all together the global currency should play a key role in reducing the costs and increasing the value of the transactions so that the system is very open for all legitimate transaction and for a very less cost.

**Trust factor**

The trust factor is very important since there is a handling of money in the whole process and also since the system would be completely online the users should trust the security of the transactions and also the business information that flows into the system. This has to be achieved by building up a strong network with advanced security features and giving the users a complete safety atmosphere and strong policies and systems that would help the users report any feedback or issues. Trust being the most impor-tant consideration it proves increasingly important to make the user realize the fact that it's more secure than any other conventional way of payments or systems that are being used right now.

**User perception for the usage of digital currency**

The user perception of the currency is one of the factors that the research focussed on. The perception of user about the global digital currency should be a positive one else it's not going to succeed. Even though the global currency is a possibility within the next decade the generation of users who use this would be the people in the internet era. So the perception factor would favour the global digital currency as the globe is moving more towards a globalisation and the citizens and becoming global in the way they thing and communication and do business with. So with this mentality it's obvious that the digital currency would be highly a user supported thing.

**ANALYSIS OF RESULTS**

**Positive Outlook**

The following questions have been answered by the subjects of the survey:

If there is a Universal currency all over the world do you think it would increase E-commerce transactions?

Can Virtual Currency really influence the way it's used without having it physical in terms of Notes and Coins?

What about loading funds to your mobile and using your mobile for every purchase? Do you think it will be easy for transactions?

In terms of the monetary size of the transactions which would you prefer doing through your mobile?

Do you think a standardized m-payment method through a universal digital currency would increase the convenience for online transactions?

Do you think Your Bank or Your Forex Dealer gives you a fair Exchange rate?

Do you think a Virtual Currency or Loyalty Points or Air Miles Should have Equivalent Real time Forex or Real life currency value?

Each one of these questions produced results that showed for a positive outlook of a universal digital currency. Questions 1-2 show the general popularity of the idea, 3-5 describe that mobile payments would be the ideal medium for the infrastructure, and 6-7 show that the subjects of the survey are unhappy with the current infrastructure of monetary exchange, and therefore would prefer a different method.

*Questions 1-2:*

The subjects of the survey showed their positive opinions of virtual currency in general through these questions. In response to whether it will increase worldwide E-commerce transactions, 73% answered yes, and 77% answered yes for virtual currency having an impact even though it will not have a physical tangible medium like notes or coins. These answers show that people are willing and ready to further their experience with online transactions, and see the benefit of doing so. Physical currency is slowly being replaced, and it is evident that the subjects understand that, and are ready to enter into a new era of currency.

*Questions 3-5:*

Many of the questions on the survey asked about using mobile devices as a medium or a method of payment. This is due to the major increase of the number of mobile devices recently. In general, the subjects of the survey responded positively to the idea of using their mobile devices for payment. Loading funds onto a mobile for ease of transactions received a 77% positive outlook. This speaks well for digital currency because the infrastructure for mobile payments already exists, which eases the implementation process a great deal.

In terms of the size of transactions that the subjects of the survey were willing to pay, an overwhelming majority of the answers were in favour of small to medium transactions (up to 1000). Most of the subject's every-day purchases are made within this payment bracket, which would give the highest volume of transactions. This uses the core competencies of the mobile gateways, as their infrastructure supports small, high volume traffic.

Lastly, most of the subjects of the survey responded positively to using m-payments online as compared to in person (89%). This question shows the unification of the payment variability, such that as long as one has their mobile device or mobile number, one can make a payment regardless of the medium. This greatly increases convince, and removes the need to carry not only notes and coins, but purses and wallets all together. The subjects of the survey see the benefit of this increased flexibility and are awaiting an infrastructure to support it.

*Question 6-7:*

Many companies have tried to take advantage of cross border transactions, or virtual currency, but have failed due to poor business models. Some questions on the survey asked about such companies, and whether the subjects would prefer something better. When asked about the exchange rates of Foreign Exchange dealers (Forex), 65% stated that they are forced into an unfair exchange rate. Air miles also came into question, and 69% of the subjects agreed that there should be some possibility of exchange for real currency. These contracts make customers lose interest in investing time and money, and reduce tourism due to high overheads of exchange rates. The subjects of the survey are looking for a better exchange rate or a universal currency so travelling becomes fiscally reasonable. This bodes well for a universal digital currency, as many people will join due to lack of interest and trust in other companies.

**Negative Outlook**

The negative outlooks can be categorized in four different aspects. The factors in essence would act as barriers and limitations to the adoption of a universal digital currency. Based on our survey and its results, the negative outlook in towards implementing a universal digital currency our categorizations are as:

- Control and Trust Factor
- Lack of Knowledge among consumers
- Service Charges
- Regulation

In terms of the questions answered by the respondent of our survey, we have identified five questions based on the answers which indicate consumer un-eagerness for a digital currency. The questions are:

1. Would let your mobile operator handle your money for online shopping acting a substitute for your bank?

2. Virtual Currency would be offered by a private entity. Do you trust private firms handling your money?

3. Do you know how much VISA, MasterCard or Paypal Charge when you use your Card for buying in a different currency other than your home country currency?

4. Do you know how much do websites, which accept payment for products and services, pay to payment processing companies?

5. Do you think the virtual currency should have no setup cost, or transaction fee involved?

We have analyzed the results in terms of the above mentioned factors:

**Control and Trust Factor:**

When it comes to trust, there seems to be a lack of it in regards to letting the mobile operator or a private entity handle it. Traditional banks are seen as the custodian of people's money and consumers are not ready to let their money change hands yet. More than half of the people said no to let their mobile operator being in charge of their money if they came up with a virtual currency. Almost the same percentage of people said no also to let any other private entity being the custodian. As our target audience was more based in locations where there is already established banking services and people have access to chip and card facilities – credit or debit cards, the need for another player stepping in this arena seemed redundant to people. But it is a proven fact in places where there was no banking service, mobile operators have successfully rendered these services to the people and are the ideal candidate for bringing the virtual currency to the people also in business viable way.

**Lack of Knowledge**

There seems to be huge gap in the knowledge and awareness about how much people actually pay as service charges currently when they are shopping online. The complex interconnectivity structure in place currently for payment gateways results in quiet a lot of service charges for consumers. The surprising thing is that they are not aware of this. About 47% people did know how much service charge they pay while using their credit cards. Most of them were also unaware how of the fees charged by the websites for their products and services went to the gateways and other payment enablers. This relevant lack of knowledge can be hindrance for creating a proper buy-in for the universal virtual currency. As we will discuss in the next section about the necessity of service charges, the current obliviousness of people would not help for creating a road map to the all digital currency.

**Service Charges**

Even though most of our survey respondents are not aware of the service charges that are in place currently, a simplified and robust model for the charges is necessary for a digital currency. This would used to provide quality in terms of service, credibility and security. But from our survey we found that majority of the respondents wouldn't like to pay a setup cost or service charge.

**Regulation**

On other point that did not directly came out of our survey but we found from our literature review is that the adoption of universal digital currency can be hindered by to much regulation. The current monetary structure regulated by banks is so rigid that it is not helpful the growth of e-commerce. But on the flipside government regulation of any monetary service is necessary for preventing money laundering and generating tax revenues for the state.

**POSSIBLE IMPLEMENTATION STRATEGIES**

In order to fully assess the feasibility of a universal digital currency, an implementation model must be created. The following are three possible implementation strategies to allow virtual currency to come to fruition.

**P2P:**

A peer to peer (P2P) implementation strategy would involve connecting each user of the virtual currency using a client program. Any time a transaction is requested; the client programs will connect and transfer the funds. This method must require a high level of security, as each transaction will be broadcasted over a public network. Bitcoin, A decentralized virtual currency has successfully proven that a P2P implementation of virtual currency is possible and even economically viable (Nakamoto, 2008). It would be possible to expand on the strategy that Bitcoin has taken to make a more globalized currency.

Some advantages of a P2P implementation strategy follow (Jansen, 2013):

- Allows for easy adoption of the service. Any merchant who wishes to adopt the service needs only to install the client, and register the client in a universal database. Since it is solely controlled via peer to peer, adoption fees and contracts are not an obstacle.

- Avoids difficulties with financial services. Financial institutions provide safe and secure transactions, however there are many obstacles to overcome. A universal approach is the primary goal of this study, and it becomes exceedingly difficult to unify the regulations to allow for this. Allowing merchants to trade without a third party makes for a more universal solution that avoids regulation problems.

There are, however some problems with a P2P solution. There is currently little infrastructure that supports a peer to peer transaction system, and the implementation process would be very gradual. Much effort is required by the individual merchants to maintain the service, and small frequency transaction customers may not even bother with the process.

**Mobile Networks:**

A mobile network implementation strategy would involve mobile devices as a medium of payment. Charges would be made to existing accounts with the mobile network. The accounts would be mapped to the device number itself rather than the device to allow purchases over e-commerce websites or in person when the device is not available (Xinhao, Jiajia and Xiufang, 2009).

The advantages of using mobile networks to implement a universal strategy are as follows:

- Uses existing infrastructure for payments

- Each mobile carrier is a supplier and manages all transactions

- The customer's phone number becomes a method of payment

- A vast percentage of the population has mobile devices

- Uses core competencies of the mobile network: Micro-transactions, security, high volume network

This implementation strategy has been proven in Japan with a company called NTT Docomo. The matured mobile market allowed for a large enough clientele bases that could utilize the payment gateway. A more globalized version of this service would be the ideal target for a universal virtual currency.

Though this seems as a better fit than a peer to peer network, there are some problems detailed below:

- Contract terms reduces flexibility

- Possible fees when changing mobile providers

- Lack of trust

Although most customers of mobile devices trust their network with their mobile activity, a different story is observed when it comes to allowing the mobile network handle financial services. The feasibility survey found that only 9% of the subjects were comfortable to use their mobile network as a substitute for a bank. This lack of trust shows that there are many obstacles that must be overcome to use mobile networks for an implementation of a universal digital currency.

**Financial Institutions:**

The last possible implementation strategy for a universal virtual currency would be to communicate directly with financial institutions. This would provide a direct link between merchants and their financial services. Each transaction would identify the client, and deposit or withdraw the funds from their account.

A successful example of this implementation method is PayPal. Merchants and consumers use PayPal as a safe and secure payment method online. Many countries like Iran and Egypt have banned PayPal due to ease of tax evasion and lowering of GDP. For this reason, a universal virtual currency must have the ability to register as a merchant in order to pay taxes to the given country.

The largest drawback with using financial institutions as a payment gateway is the sheer volume of rules and regulations, along with the necessary exchange rate deductible when transferring funds between countries. Most financial institutions are controlled by the government, and enforce laws based on the idealisms of the country. Also, when using concrete currency as a medium, it is unavoidable to pay a percentage when transferring funds between countries. That means that a Forex dealer must be involved, which adds another layer of complexity to the implementation process.

In conclusion, each payment gateway has their advantages and their drawbacks. The demand for a universal digital currency is strong, but existing payment infrastructure may need a more unified approach before lossless transactions can be made. Both the feasibility survey and the analysis of the implementation strategies point to a mobile payment gateway as the best medium, but perhaps other opportunities could arise in the future.

**SECURITY**

When managing monetary transactions, security becomes the upmost priority. Most of the foreseeable transactions will occur over a public network, where any attacker could intercept a transaction. The makers of Bitcoin used a multi-tier approach to encrypt each transaction, which makes decryption of the message virtually impossible. There are other ways, however, to maliciously tamper with virtual currency.

Since the implementation method of universal digital currency is currently unforeseeable, it is difficult to discuss how the medium of a transaction can be protected from malicious attackers on the on public networks. However, identity theft also poses a problem. With each implementation method discussed above, it would be possible that theft of a device or wallet could allow for erroneous and malicious use of the currency.

Biometrics would help ensure that the consumer's funds are only transferrable when identified. This field has gone through several stages of maturity (Jain, Hong and Pankanti, 2000), and the feasibility study shows that the subjects are willing to secure funds using a biometric security method. Using biometrics as a payment gateway is a proven business strategy. Amusements parks are allowing customers to use their fingerprint as a payment method. When entering the park, the consumer scans their finger, and is asked if they would like to load funds onto their account. Later, when making a purchase, the customer can swipe their finger, and the funds will be taken off of their account.

This kind of technology shows large promise when coupled with universal digital currency. Users will be able to trust that their funds are safe, as biometrics ensures that an identity match must be made to make a transaction. It is the maturity of technology like this that will allow for a larger clientele base for a universal digital currency, and therefore aide its feasibility.

## LIMITATIONS OF OUR SURVEY

We tried to get our survey out to as much as people. Despite of that our survey was limited in terms of the variety of the respondents. Due to a lack of time and access the there was relevant shortage of industry personnel and young people. These are some areas that can be looked into in regards to future work in this field.

## CONCLUSION

The need for a global currency is the need for the hour. But the feasibility seems to be very low looking at the way the co-ordination of different stakeholders in the world including different regulatory bodies and countries involved in the political and economical spectrum. Though there are sound reasons to believe that this may happen in the distant future looking at the overall benefits for each other in the system this seems impossible in the immediate future. The impact of such Global currency for e-commerce would be enormous in terms of ease of transaction and also reaping the benefits of all the potential of the businesses. The evolving community currency structures could be a great solution for this but again the community currencies would still face all the regulatory issues and acceptance hurdles. New currency systems should try to address all the notes issues from the survey results and bring in new solutions for the existing issues users have to gain wider usage and acceptance.

**APPENDIX: SURVEY RESULTS**

1. If there is a Universal currency all over the world do you think it would increase E-commerce transactions?

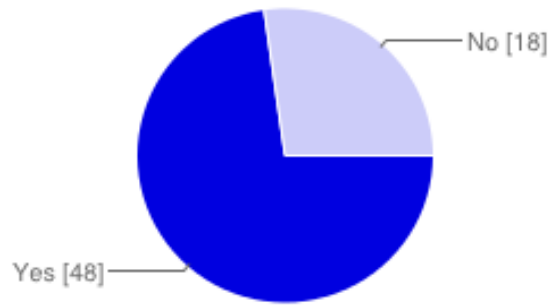

| | | |
|---|---|---|
| Yes | **48** | 73% |
| No | **18** | 27% |

2. Do you think the use of printed currency would decrease in the next few years?

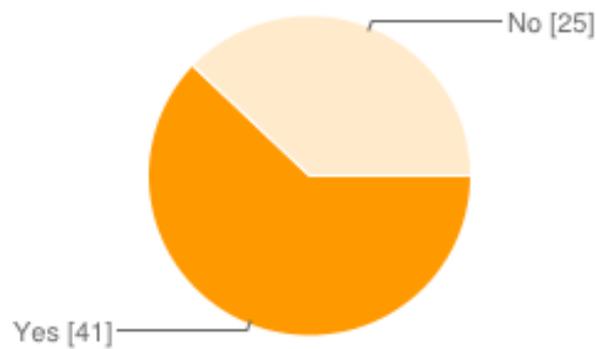

| | | |
|---|---|---|
| Yes | **41** | 62% |
| No | **25** | 38% |

3. Would you support a Bio-metric payment system?

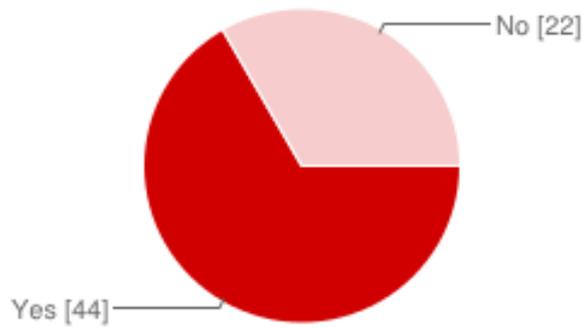

| | | |
|---|---|---|
| Yes | **44** | 67% |
| No | **22** | 33% |

4. What about loading funds to your mobile and using your mobile for every purchase? Do you think it will be easy for transactions?

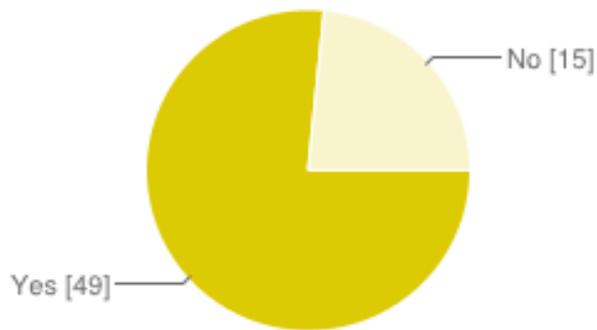

| | | |
|---|---|---|
| Yes | **49** | 77% |
| No | **15** | 23% |

5. In terms of the monetary size of the transactions which would you prefer doing through mobile?

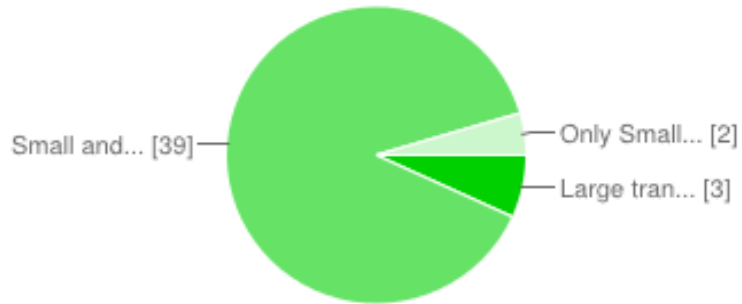

Large transactions (above $1000)

Small and Medium transactions (Up to $1000)

Only Small transactions (Up to $30)

6. Would you let your mobile operator handle your money for online shopping acting a substitute for your bank?

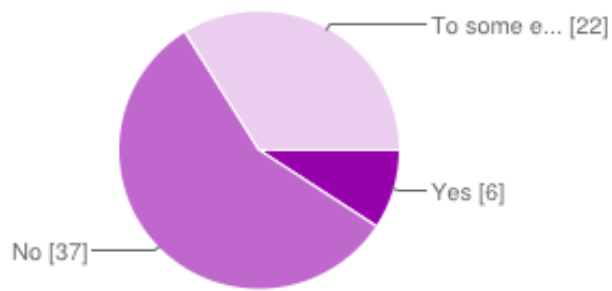

| | | |
|---|---|---|
| Yes | **6** | 9% |
| No | **37** | 57% |
| To some extent | **22** | 34% |

7. When you spend online using net banking or Credit card its widely said that you don't feel the currency value but when you take the cash physically out from your wallet you feel the value more. Would this make people insecure with virtual currencies on a longer run?

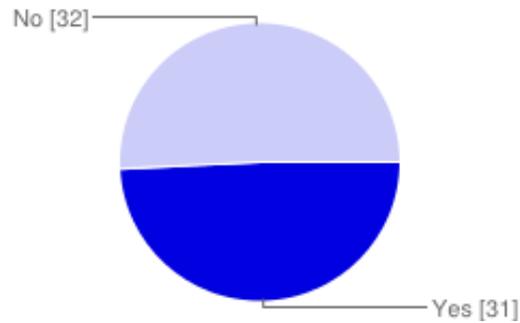

| | | |
|---|---|---|
| Yes | **31** | 49% |
| No | **32** | 51% |

8. Do you think the virtual currency should have no setup cost, or transaction fee involved?

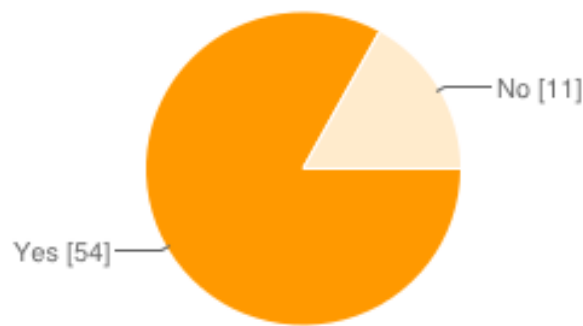

| | | |
|---|---|---|
| Yes | **54** | 83% |
| No | **11** | 17% |

9. Do you know how much VISA, MasterCard or Paypal Charge when you use your Card for buying in a different currency other than your home country currency?

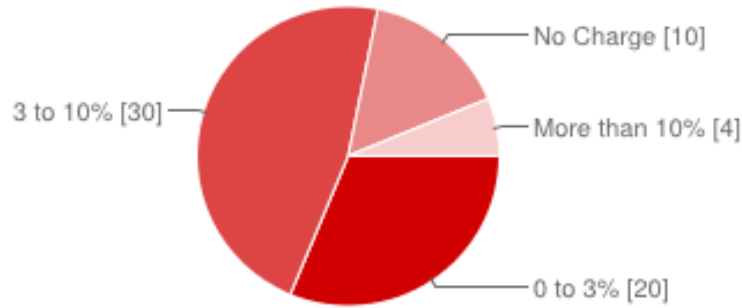

| | | |
|---|---|---|
| 0 to 3% | **20** | 31% |
| 3 to 10% | **30** | 47% |
| No Charge | **10** | 16% |
| More than 10% | **4** | 6% |

10. Do you know how much do websites, which accept payment for products and services, pay to payment processing companies?

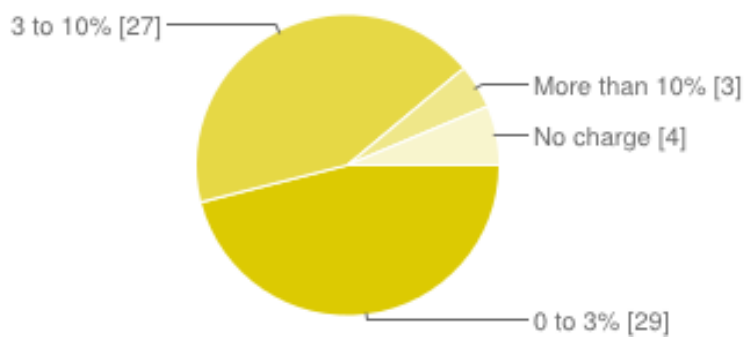

| | | |
|---|---|---|
| 0 to 3% | **29** | 46% |
| 3 to 10% | **27** | 43% |
| More than 10% | **3** | 5% |
| No charge | **4** | 6% |

11. What do you think is the best way to redeem the virtual currency?

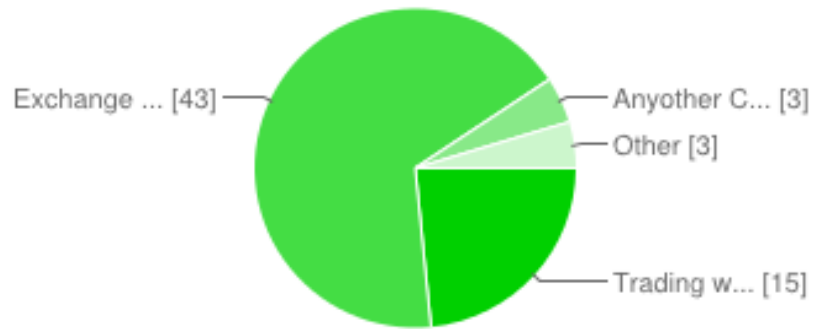

| | | |
|---|---|---|
| Trading with Buyers and Seller | **15** | 23% |
| Exchange rate with real life currency | **43** | 67% |
| Any other Credits/ Loyalty points etc | **3** | 5% |
| Other | **3** | 5% |

12. Do you think Your Bank or Your Forex Dealer gives you a fair Exchange rate?

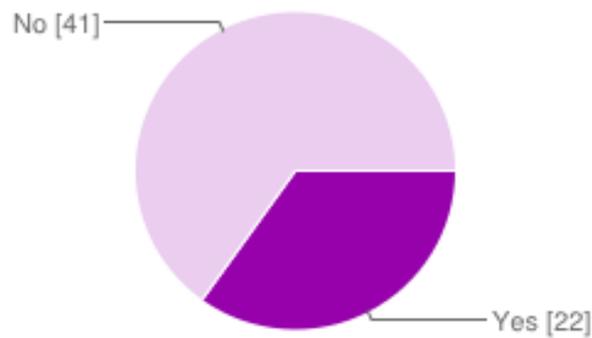

| | | |
|---|---|---|
| Yes | **22** | 35% |
| No | **41** | 65% |

13. Do you think a Virtual Currency or Loyalty Points or Air Miles Should have Equivalent Real time Forex or Real life currency value?

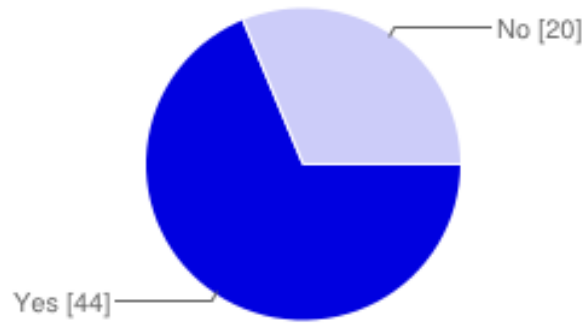

| | | |
|---|---|---|
| Yes | **44** | 69% |
| No | **20** | 31% |

14. Whats the commission or amount do Forex dealers charge?

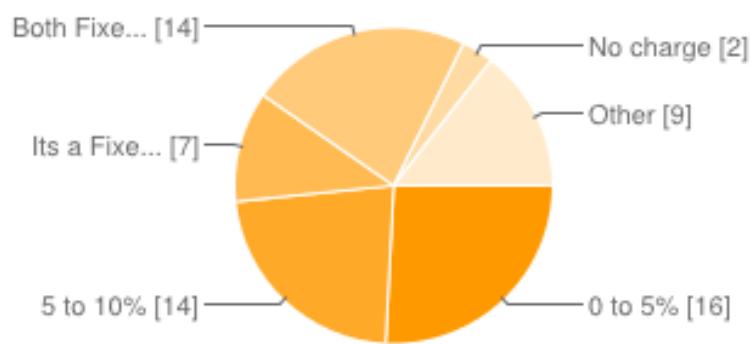

| | | |
|---|---|---|
| 0 to 5% | **16** | 26% |
| 5 to 10% | **14** | 23% |
| Its a Fixed fee | **7** | 11% |
| Both Fixed and Percentage | **14** | 23% |
| No charge | **2** | 3% |
| Other | **9** | 15% |

15. Can Virtual Currency really influence the way its used without having it physical in terms of Notes and Coins?

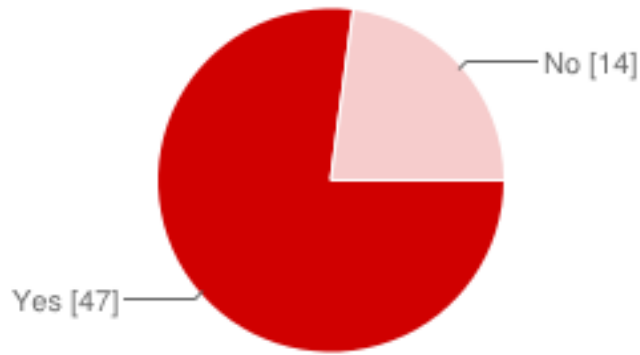

| | | |
|---|---|---|
| Yes | **47** | 77% |
| No | **14** | 23% |

16. Virtual Currency would be offered by a private entity. Do you trust private firms handling your money?

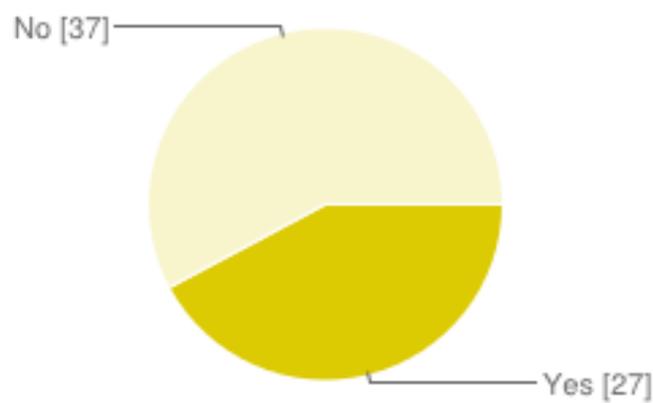

| | | |
|---|---|---|
| Yes | **27** | 42% |
| No | **37** | 58% |

17. If everyone starts using virtual money do you think it would make governments insecure for Tax collection, Money laundering?

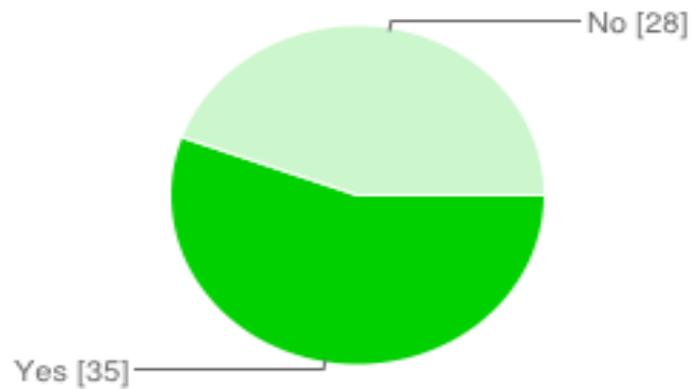

| | | |
|---|---|---|
| Yes | **35** | 56% |
| No | **28** | 44% |

18. Do you think virtual money would have economic implications like the money would not count towards GDP or inflation etc?

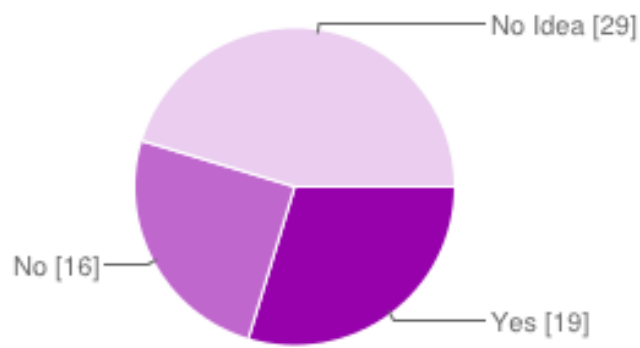

| | | |
|---|---|---|
| Yes | **19** | 30% |
| No | **16** | 25% |
| No Idea | **29** | 45% |

19. Would you like the capability of sending virtual money to friends or relatives through Facebook?

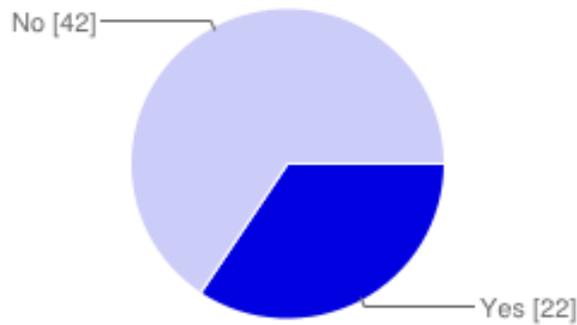

| | | |
|---|---|---|
| Yes | **22** | 34% |
| No | **42** | 66% |

20. Do you think a standardized m-payment method through a universal digital currency would increase the convenience for online transactions?

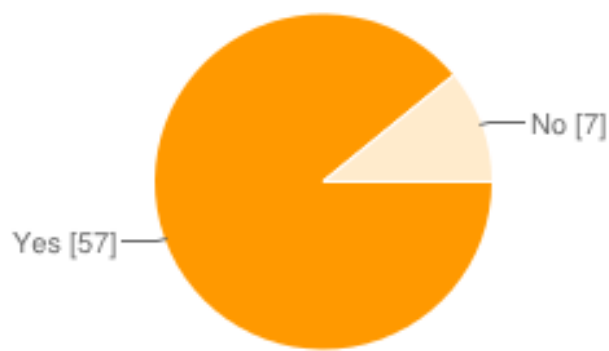

| | | |
|---|---|---|
| Yes | **57** | 89% |
| No | **7** | 11% |